\documentclass[prd,twocolumn,showpacs,preprintnumbers,nofootinbib,amsmath,amssymb,nobalancelastpage]{revtex4}
\usepackage{graphicx}
\usepackage{bm}
\usepackage{dcolumn}
\usepackage{amsmath}
\usepackage{amssymb}
\usepackage{color}
\usepackage{url}
\usepackage{hyperref}

\usepackage{aas_macros}
\usepackage[normalem]{ulem}

\def\Eq#1{Eq.~(\ref{#1})}

\def\cm{\rm{cm}}
\def\g{\rm{g}}
\def\erf{\rm{erf}}
\def\km{\rm{km}}
\def\s{\rm{s}}
\def\GeV{\rm{GeV}}
\begin{document}

\title{Self-interacting dark matter constraints in a thick dark disk scenario }

\author{Kyriakos Vattis}
\email{kyriakos\_vattis@brown.edu}
\author{Savvas M. Koushiappas}
\email{koushiappas@brown.edu}
\affiliation{Department of Physics, Brown University, 182 Hope St., Providence, Rhode Island 02912, USA}

\date{\today}

\begin{abstract}
A thick dark matter disk is predicted in cold dark matter simulations as the outcome of the interaction between accreted satellites and the stellar disk in Milky Way sized halos. We study the effects of a self-interacting thick dark disk on the energetic neutrino flux from the Sun.  We find that for particle masses between 100 GeV and 1 TeV  and dark matter annihilation to heavy leptons either the self-interaction may not be strong enough to solve the small scale structure motivation  or a dark disk cannot be present in the Milky Way.
\end{abstract}


\maketitle
 
\section{\label{sec:Intro}Introduction}
In the last two decades a cosmological model emerged with the Universe being composed of Cold Dark Matter (CDM) and a dark energy with a negative equation of state. In this cosmology, structure forms hierarchically, small scales form first and their subsequent mergers lead to the formation of larger scale objects up to the scales of galaxy clusters. Adiabatic perturbations in the early universe seed the formation of structure and lead to the development of voids and filaments of dark matter. In this scenario, the formation of a large object follows the accretion of smaller systems along filaments that provide the mass funneling to the central deep potential well of dark matter halos. 

The orbital evolution of accreted satellites is heavily influenced by the presence of the Galactic disc. The result is that the stars and gas drag the accreted satellites, tidal forces destroy these satellites and their dark matter settles into a dark thick co-rotating disk that is co-planar to the Galactic disk \cite{read2008thin,2009MNRAS.397...44R}. This thick dark disk is a stable configuration of dark matter in phase space unlike dark matter streams that disperse in timescales set by the velocity dispersion of dark matter in the host halo. The existence of such a disk has implications in experimental attempts for dark matter detection  \cite{bruch2009dark}.

The nature of dark matter interactions among itself and among standard model particles remains unknown. Searches for Weakly Interacting Massive Particle (WIMP) signatures have at most given rise to suspicious hints with no clean significant detection \cite{2017Natur.552...63D,2015PhRvL.115h1101G}, and the parameter space for compact halo objects as a dark matter candidate is also shrinking \cite{2017PhRvL.119d1102K,2017ApJ...838....8L}. Other dark matter candidates are in the form of an axion-like particle \cite{schive2014cosmic} or a particle with a self-interaction cross section \cite{spergel2000observational}. Both of these candidates have interesting consequences to a plethora of problems within the context of CDM \cite{de2010core,boylan2011too,boylan2012milky}
 
Self interacting dark matter particles were tested by analysis of the matter distribution of the Bullet Cluster as well as  galaxy cluster shapes  excluding the parameter space of $\sigma_{\chi\chi}/m_\chi > 1.25 \, \cm^2 / \g$ \cite{randall2008dark} and $\sigma_{\chi\chi}/m_x > 1.0 \, \cm^2/\g$ \cite{peter2013cosmological,rocha2013cosmological} respectively.  It should be noted that the limits obtained from the bullet cluster do not suffer from the systematics inherent to the modeling assumptions of the shape of galaxy clusters, and thus the bullet cluster may be considered the cleanest test of self-interaction from these two observations. 

If the dark matter particle has a non-zero self-interacting cross section then the capture rate of such particles in the Sun and the Earth \cite{gould1987resonant,gould1992cosmological} will be enhanced compared to the standard approach  where the capture rate is simply set by the scattering cross section of the dark matter particle with standard model particles \cite{zentner2009high}. IceCube placed constraints on self-interacting dark matter by looking at the energetic neutrino flux from dark matter annihilation in the center of the Sun. In addition, \citet{albuquerque2014constraints} independently probed the same parameter space confirming most of the excluded space from the previous analysis. 

In this manuscript we calculate constraints to the self-interacting cross section of dark matter in two-scenarios: First we use the latest results from IceCube \cite{aartsen2017search} and compute updated limits for annihilation to heavy leptons ($\tau^+\tau^-$) and heavy quarks ($b\bar{b}$) in the case of the Standard Halo Model (the case of annihilation to gauge bosons gives limits that are in between the two cases we consider here). This is similar in spirit to the analysis of \citet{albuquerque2014constraints}. We then repeat the calculations under the assumption that a thick dark matter disk alters the density and velocity distribution of dark matter at the solar radius. The importance of this second calculation is in the fact that since the dark disk is very likely to be co-rotating with the visible disk of the galaxy or to have a small lag speed there is a significant enhancement to the predicted capture rate and subsequently the annihilation rate. 

We begin in Section II where we show how dark matter capture and annihilation rates are determined by the density and velocity distribution of dark matter and how these rates are affect by the presence of a thick dark disk at the solar neighborhood. Our results are presented in Section III, and we conclude in Section IV. 

\section{\label{sec:Intro}Dark matter capture and annihilation rates}
In most scenarios captured dark matter is thermalized in the solar interior on a timescale that is less than the age of the Sun ($\tau_{\odot}=5 \times10^9$ yr) and the time evolution of the number of captured particles $N_\chi$ is given by
\begin{align}
\label{eq:dmpnumber}
\frac{dN_\chi}{dt}=C_c+C_sN_\chi-C_aN_\chi^2,
\end{align}
where $C_c$ is the capture rate of dark matter by interactions with nuclei in the Solar core, $C_sN_\chi$ is  the capture rate by dark matter self-interactions, and $C_aN_\chi^2$ twice the annihilation rate per pair of dark matter particles \cite{zentner2009high}. We ignore the effects of time dependence that can arise in the case of rich substructure in phase space \cite{2009PhRvL.103l1301K}.

The general solution of \Eq{eq:dmpnumber} is
\begin{align}
\label{eq:dmpnumbersol}
N_\chi=\frac{ C_c\ \tanh( t / \zeta )}{ \zeta^{ -1 } - C_s\  \tanh( t / \zeta ) / 2 }
\end{align}
where
\begin{align}
\label{eq:zeta}
\zeta=\frac{1}{\sqrt{C_cC_a+C_s^2/4}}. 
\end{align}
The quantity $\zeta$ is the equilibration timescale, which is the timescale where the rate of capture of dark matter (from scattering with nuclei and itself) equals the annihilation rate of dark matter particles. 

By knowing the number of  captured particles $N_\chi$ at the time equal to the current age of the Sun $\tau_{\odot}$ the annihilation rate can be calculated as
\begin{align}
\label{eq:anhrate}
\Gamma_a=\frac{1}{2}C_aN_\chi^{2}.
\end{align}
It is interesting to notice that in the limit of no self interactions and at times much greater than the equilibration timescale the annihilation rate reduces to 
\begin{align}
\label{eq:anhratens}
\Gamma_a=\frac{1}{2}C_c.
\end{align}
On the other hand when self interactions are dominating the capture, i.e., $C_s^2 \gg C_c C_a$,  the annihilation rate is 
\begin{align}
\label{eq:anhrateds}
\Gamma_a =\frac{1}{2} \frac{C_s^2}{C_a}, 
\end{align}
independent of $C_c$.

The annihilation rate coefficient $C_a$ depends on the dark matter distribution in the Sun and is usually given in terms of the effective volumes $V_j$ \cite{gould1987resonant,jungman1996supersymmetric,zentner2009high} by 
\begin{align}
\label{eq:anniltratecoef}
C_a=\left<\sigma_A v \right>\frac{V_2}{V_1^2}
\end{align}
where
\begin{align}
\label{eq:effvol}
V_j=2.45 \times 10^{27}\left(\frac{100 \, \GeV}{ j \, m_{\chi} } \right)^{3/2} \cm^3
\end{align}
and $\left<\sigma_A v \right>$ is the velocity averaged annihilation cross section which for a non-relativistic thermal relic dark matter particle is roughly $3\times10^{-26} \cm^3 / \s$ \cite{2012PhRvD..86b3506S}.

We assume that the spin-independent cross section for dark matter scattering off nuclei heavier than Hydrogen is given by \cite{zentner2009high} 
\begin{equation}
\label{eq:sics}
\sigma^{SI}_{\chi n} = \sigma^{SI}_{\chi p} \, A^2  \left( \frac{m_n}{m_p} \right)^2 \frac{(m_\chi+m_{p})^2}{(m_\chi+m_n)^2} \\
\end{equation}  
where $A$ is the atomic mass number, $m_{p}$ is the proton mass and $m_n$ is the mass of the nucleus.

The dark matter capture rate due to elastic scattering off  an element distributed in the Sun $C_c$ is (see   \citet{gould1987resonant,gould1992cosmological,jungman1996supersymmetric}) 
\begin{widetext}
\begin{equation}
\label{eq:nucleirate}
\begin{split}
&C_c=\frac{ \rho_x \, M_\odot \, v_\ast \, f \sigma_{\chi n}}{ a \, m_\chi \, m_n \, \beta_+} \left[ \frac{2 \, \exp(- a \hat{\eta}^2)}{(1+ a )^{1/2}} \, \erf(\hat{\eta}) - \frac{ \exp(- a \hat{\eta}^2)}{(A^2_c-A^2_s)(1+ a)^{3/2}}\times\bigg\{\left(\hat{A}_+\hat{A}_- -\frac{1}{2}-\frac{1+a }{a-b}\right)[\erf(\hat{A}_+)-\erf(\hat{A}_-)]\right. \\
&+ \frac{1}{\sqrt{\pi}}  \left( \hat{A}_-e^{-\hat{A}_+^2}-\hat{A}_+e^{-\hat{A}_-^2} \right) \bigg\}^{A_c}_{A_s} +\frac{\exp(-b\check{\eta}^2)}{(a-b)(A_c^2-A_s^2)(1+b)^{1/2}}\{[2 \,\erf(\check{\eta})-\erf(\check{A}_+)+\erf(\check{A}_-)]e^{-(a-b)^2A^2}\}^{A_c}_{A_s} \bigg]
\end{split}
\end{equation}
where 
\begin{equation}
\label{eq:nucleiratenot}
\begin{split}
\nonumber
&v_\ast=\sqrt{\frac{2}{3}} \frac{\bar{v}}{\eta} , \ \ \ a=\frac{m_\chi v_\ast^2}{2E}, \ \ \ b=\beta_+a, \ \ \  \hat{\eta}=\frac{\eta}{(1+a)^{1/2}},\ \ \ \check{\eta}=\frac{\eta}{(1+b)^{1/2}}, \ \ \ \hat{A}=A(1+a)^{1/2}, \ \ \ \check{A}=A(1+b)^{1/2}  \\
&\hat{A}_{\pm}=\hat{A}\pm\hat{\eta}, \ \ \ \check{A}_{\pm}=\check{A}\pm\check{\eta},  \ \ \ A^2(v)=\beta_-\frac{v^2}{\bar{v}^2}, \ \ \ A_c=A(v_c), \ \ \  A_s=A(v_s) \ \ \ {\rm{and}} \ \ \  \beta_\pm=\frac{4m_nm_\chi}{(m_\chi\pm m_n)^2}.
\end{split}
\end{equation}
\end{widetext}
In this pedagogical expression, $\rho_x$ is the dark matter density,  $m_\chi$ is the mass of the dark matter particle, and $M_\odot$ is the mass of the Sun. 
$\sigma_{\chi n}$ is the elastic scattering cross section between dark matter and a nucleon, and we assume a Maxwell-Boltzman distribution for dark matter particles with a velocity dispersion $\bar{v}$. The speed of the Sun through the dark matter halo is $v_\odot$ which can be written in a dimensionless form as $\eta^2=3(v_{\odot}/\bar{v})^2/2$.

To get Eq.~\ref{eq:nucleirate} we assume that the local escape velocity of the Sun can be approximated by 
\begin{align}
\label{eq:escvel}
v^2(r)=v^2_c-\frac{M(r)}{M_\odot}(v^2_c-v^2_s)
\end{align}
where $v_c=1354 \, \km / \s$ and $v_s=795 \, \km / \s$. Note that these velocities are approximate and are based on the modeling of the Solar interior (see \citet{gould1992cosmological}) -- if one assumes the true escape velocity the effect on the scattering cross section is negligible. 
Another assumption involved in this calculation is that the form factor that parametrizes a suppression of the interaction which arises when momentum transfer if of order or greater than the inverse nuclear radius is given by an exponential of the form
\begin{equation}
\label{eq:ffactor}
F(\Delta E)=\exp \left( -\frac{\Delta E}{2E} \right),
\end{equation}
where $E=1.5 \hbar^2 / 2 m_n R^2_n $ and the mean-square radius of the nucleus $R_n$ is approximated by $R_n=[0.91(m_n/\GeV)^{1/3}+0.3]\times 10^{-13} \cm$.  

In order to get the total capture rate by nuclei in the Sun, Eq.~\ref{eq:nucleirate} needs to be summed over all elements of interest. Here, we will include the 16 most abundant elements in the Sun between Hydrogen and Nickel as in \citet{grevesse2011chemical}.

Finally the rate of self-capture is given by \citet{zentner2009high}
\begin{align}
\label{eq:selfintrate}
C_s=\sqrt{\frac{3}{2}}n_\chi \sigma_{\chi\chi}\frac{v^2_{\rm{esc}}(R_{\bigodot})}{\bar{v}} \langle \hat\phi_\chi \rangle \frac{\erf(\eta)}{\eta} 
\end{align}
and depends on the local number density of the dark matter $n_\chi$, the dark matter self interaction cross section $\sigma_{\chi\chi}$, the real escape velocity at the surface of the Sun $v^2_{\rm{esc}}(R_{\bigodot})=618 \, \km / \s$, the dimensionless  speed of Sun $\eta$  and the average potential $ \langle \hat\phi_\chi \rangle=5.14$ which accounts for the dark matter distribution (we ignore evaporation effects as they are negligible for the dark matter particle masses we consider here).  

In order to obtain limits on the strength of dark matter self interaction, we compute the annihilation rate in the Sun using Eqs. \ref{eq:anhrate} \& \ref{eq:dmpnumbersol}. We set upper limits on the self-interaction cross section $\sigma_{\chi\chi}$ and the dark matter nucleon cross section  $\sigma_{\chi n}$ by comparing the derived annihilation rate with the upper limit as measured by IceCube \cite{aartsen2017search}. 

We compute the annihilation rate  in two scenarios. First we assume the Standard Halo Model (SHM) with a Maxwell-Bolztman distribution with $v_\odot=220 \, \km / \s$ and $\bar{v}=270 \, \km / \s$ \cite{2002ApJ...573..597K} while in the second scenario we add to the SHM a thick dark disk with density $\rho_{DD}$ that is $\sim \ 0.25 \ - \ 1.5$ times the local density of dark matter in the SHM   $\rho_\chi=0.4 \GeV / \cm^3$ \cite{salucci2010dark}, a rotation lag with respect to the stellar disk of $v_\odot\sim [0 - 150] \, \km / \s$ and  near isotropic 1-D dispersion of $\bar{v}\sim [85 - 155]\, \km / \s$. 

\begin{figure}
\includegraphics[scale = 0.35]{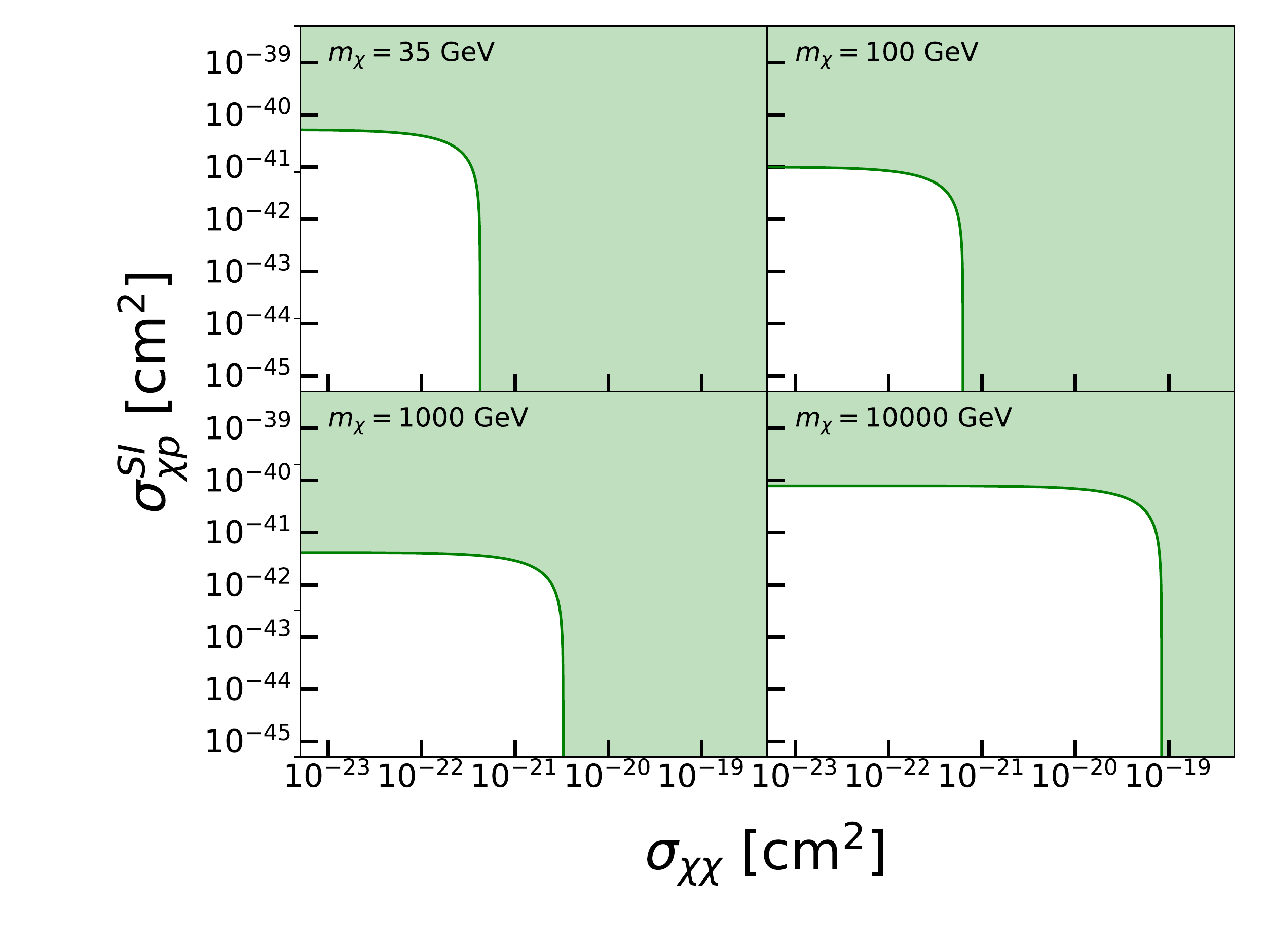}
\caption{An example of upper limit of the annihilation rate $\chi\chi\rightarrow b\bar{b}$ as set in \cite{aartsen2017search}, for various dark matter masses on the $(\sigma_{\chi p},\sigma_{\chi\chi})$ plane.  The parameter space above and to the right of each curve is excluded (confirming the behavior of \Eq{eq:anhratens} and \Eq{eq:anhrateds} in the limit where self-interaction dominates over scattering off nuclei and vice versa). }
\label{fig:bb}
\end{figure}

\begin{figure*}
\includegraphics[scale=0.35]{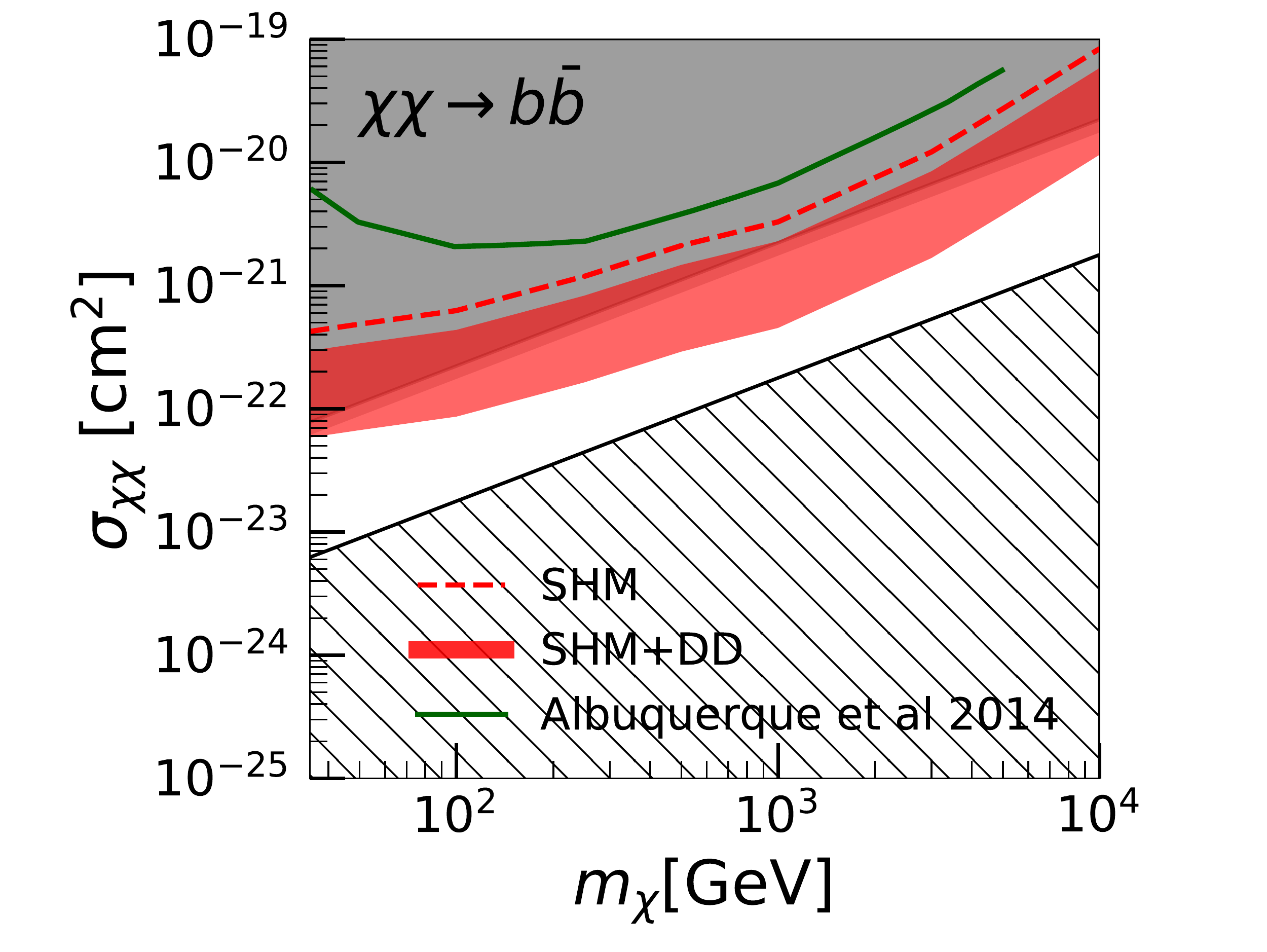}\includegraphics[scale=0.35]{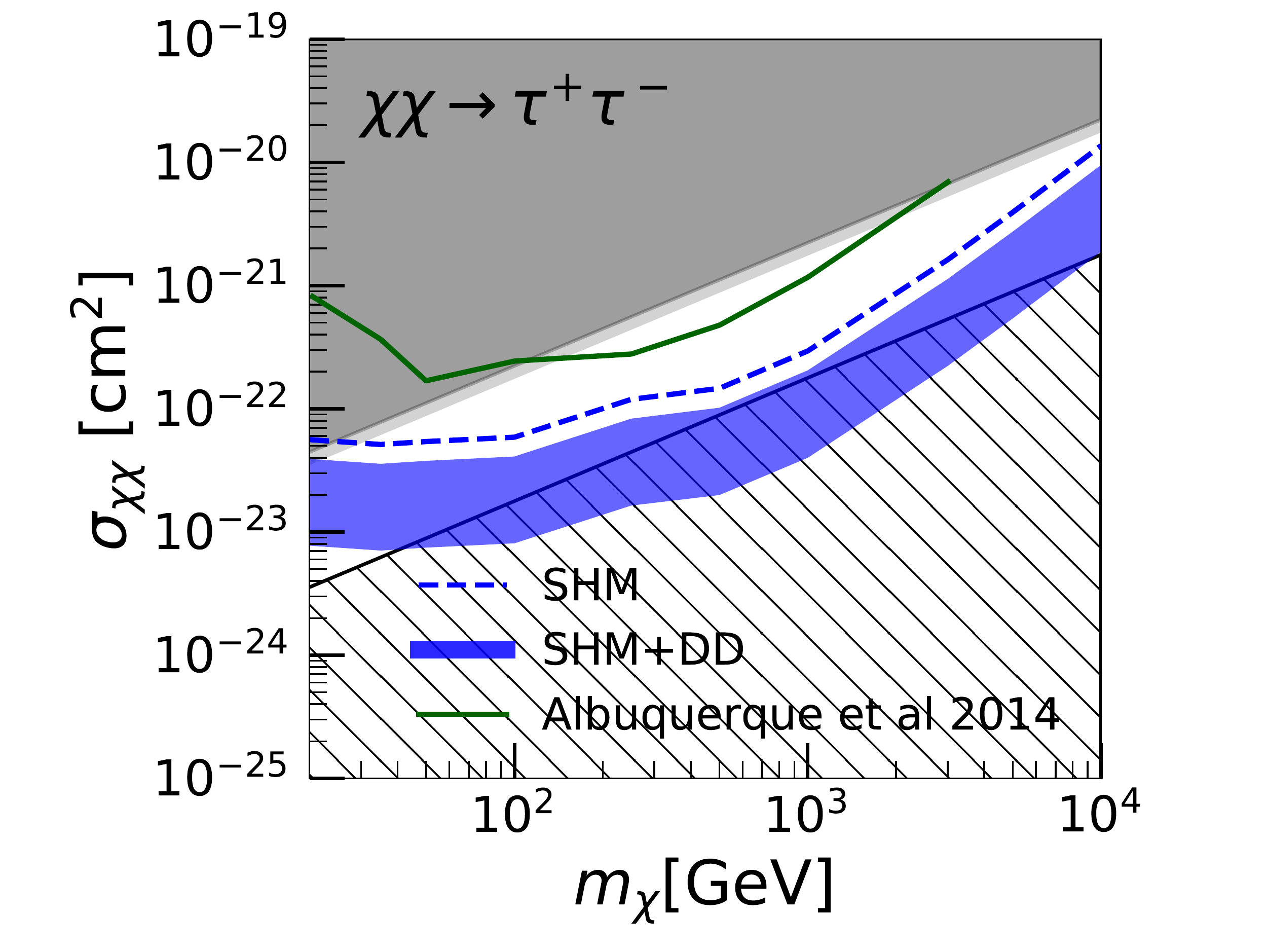}
\caption{Upper limits of the annihilation rate of dark matter on the $(\sigma_{\chi\chi},m_\chi)$ plane for annihilations to $b\bar{b}$ and  $\tau^+\tau^-$. The dashed line represents the Standard Halo Model (SHM) while the green line is the result of \citet{albuquerque2014constraints}. The colored filled regions represent the effects of the addition of a dark disk to the Milky Way halo. The upper edge of each filled region is a fast rotating disk while the bottom edge corresponds to a slow rotating disk (see text for details). The dark grey region is excluded by analysis of the Bullet Cluster \cite{randall2008dark}, the light grey region by galaxy cluster shapes \cite{rocha2013cosmological,peter2013cosmological} and the hatched area is the region of the parameter space for which the self-interaction is not strong enough to resolve the small scale structure problems \cite{zavala2013constraining}.}
\label{fig:siconst}
\end{figure*}

\section{\label{sec:Intro}Results}

We first consider self interacting dark matter without the presence of a dark disk (similar to the analysis of \citet{albuquerque2014constraints}). Figure~\ref{fig:bb} shows contours of the observed annihilation rate from IceCube in the $(\sigma_{\chi p},\sigma_{\chi\chi})$ plane assuming spin independent interaction with the nuclei and annihilation of the dark matter particles in the heavy quark ($b\bar{b}$) channel. The region above and to the right of each curve is excluded for each mass, as the rate of neutrinos (from annihilation in the solar interior) would violate the observed upper limit of IceCube \cite{aartsen2017search}. 
The sharp cutoff on each of $\sigma_{\chi p}$ and $\sigma_{\chi\chi}$ originate from equilibration considerations and the relative strength between the capture rate off nuclei and the self-interacting capture rate (see Eqs.~\ref{eq:anhratens} \& \ref{eq:anhrateds}). 

\begin{figure*}
\includegraphics[scale=0.35]{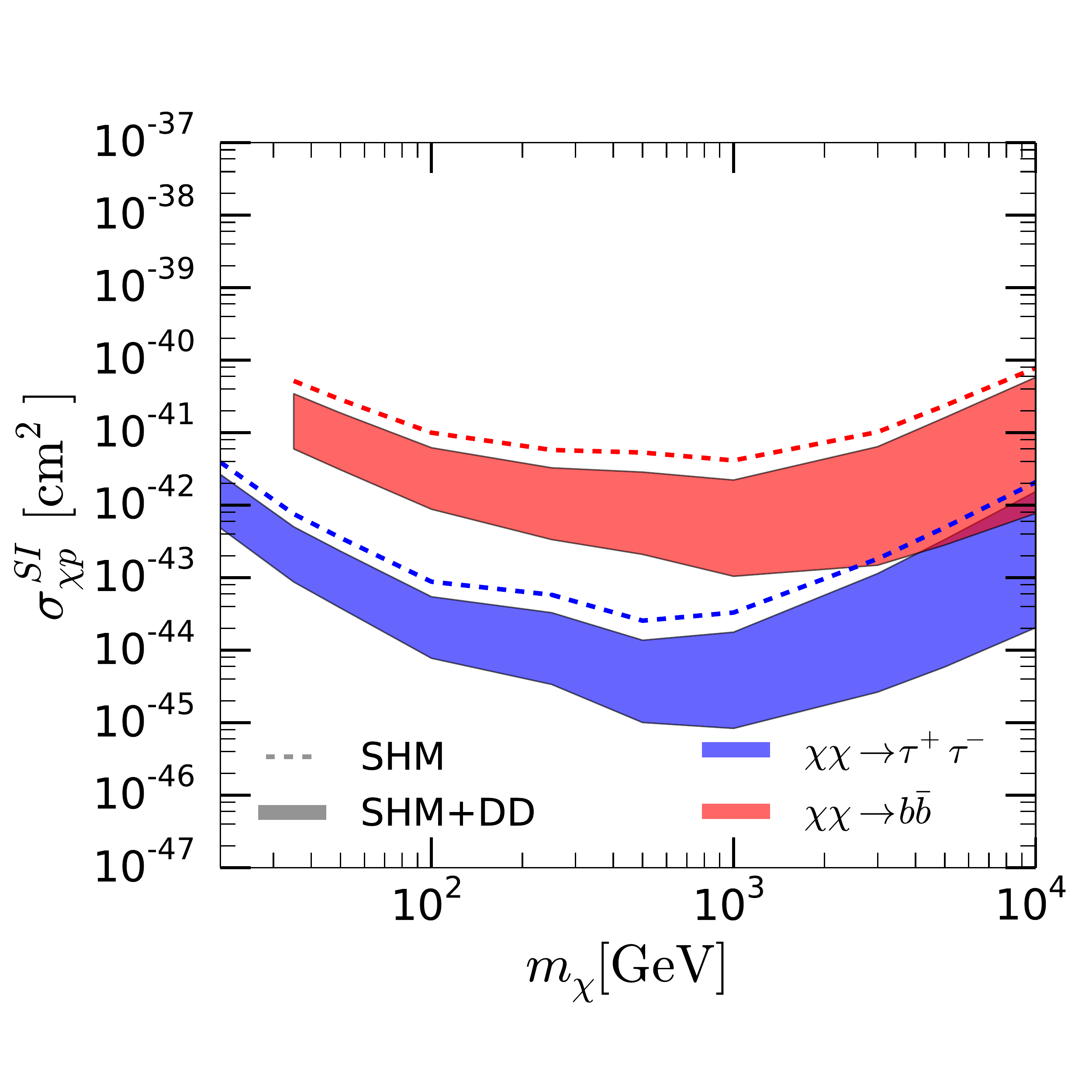}\includegraphics[scale=0.35]{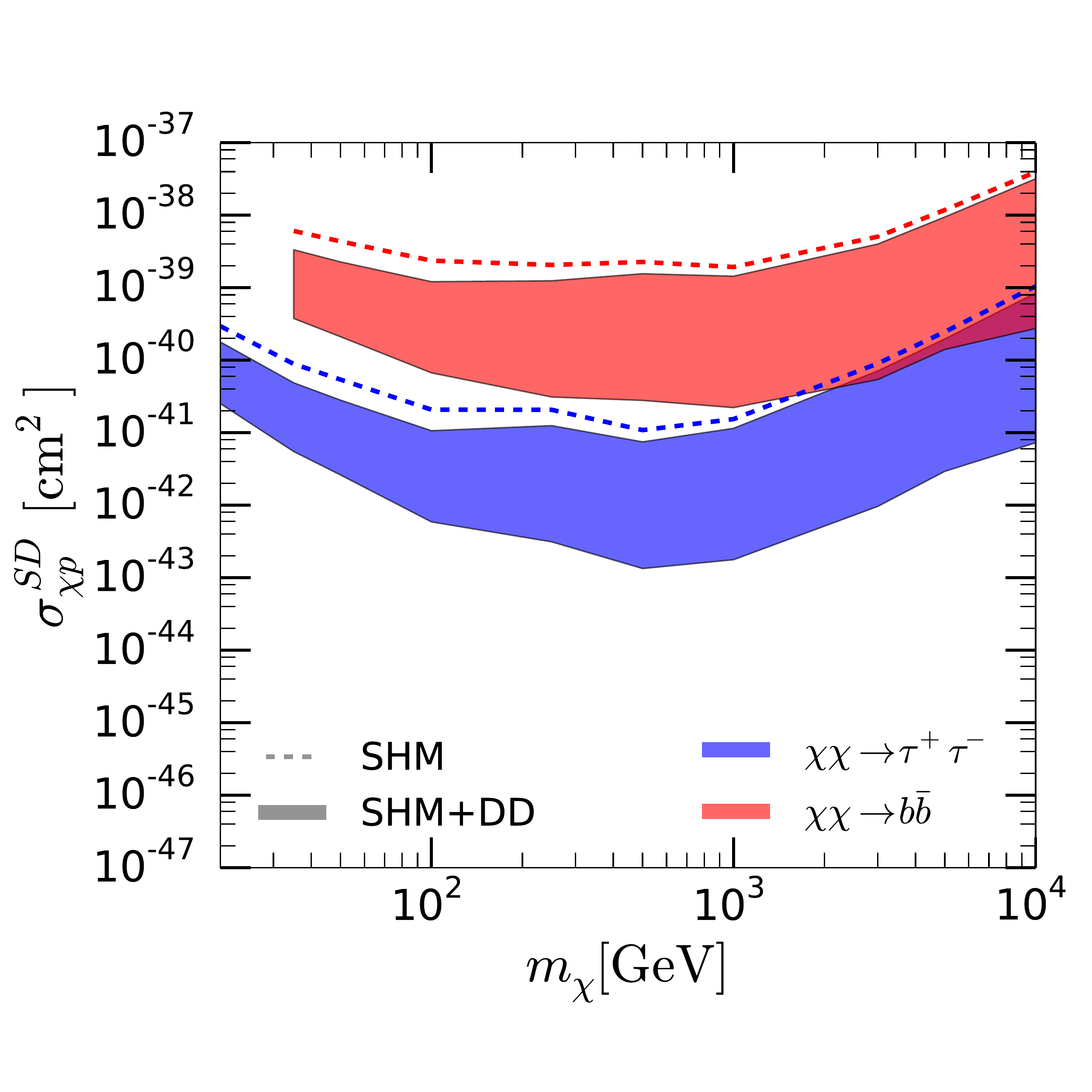}
\caption{Upper limits on the dark matter-proton scattering cross section as a function of dark matter mass for spin-independent (left) and spin-dependent (right) interactions.  The dashed line represents the Standard Halo Model (SHM) while the colored shaded areas show the limits derived with the addition of a dark disk The upper edge of each filled region is a fast rotating disk while the bottom edge corresponds to a slow rotating disk (see text for details). Note that here we assume a SHM with $\rho_\chi = 0.4 \, {\rm{GeV}}/{\rm{cm}}^3$ while the limits presented in IceCube are derived using $\rho_\chi = 0.3 \, {\rm{GeV}}/{\rm{cm}}^3$.}
\label{fig:sisd}
\end{figure*}

We next calculate constraints on the $(\sigma_{\chi\chi},m_\chi)$ parameter space using IceCube limits on the annihilation rate \cite{aartsen2017search}. We assume conservatively that $\sigma_{\chi p}=10^{-47}\cm^2$, a value well within the region that the annihilation rate is independent of $\sigma_{\chi p}$ for all the scenarios we consider as well as below the current sensitivity of direct detection experiments \cite{2017PhRvL118b1303A}. 

Figure~\ref{fig:siconst} shows the constraints on $\sigma_{\chi\chi}$ as a function of $m_\chi$ for annihilating to heavy quarks and heavy leptons. The upper limits are independent  of whether the interaction with the nuclei is Spin-Dependent (SD) or Spin-Independent (SI) because the chosen value of $\sigma_{\chi p}$ is low enough and thus the capture off nuclei is significantly smaller than self-capture (a choice of a higher value of $\sigma_{\chi  p}$ would make the limits even stronger). For the SHM model the constraints improve by a factor of $\sim2-4$ for large masses compared to \citet{albuquerque2014constraints}, as expected from the improved annihilation rate limits set by IceCube\footnote{Note  that the constrains shown in the $\tau^+\tau^-$ panel from \cite{albuquerque2014constraints}  are a combination of the  $W^+W^-$ and $\tau^+\tau^-$ channel for masses below the mass of W.}. 
The heavy quark  channel being the softest produces lower energy neutrinos and the limit is less constraining compared to annihilating in heavy lepton final states. In both cases, the limits are similar or stronger compared to the limits set by the bullet cluster for dark matter masses greater than $\sim 100$ GeV. For completeness we repeat the calculations for a gauge boson final state ($W^+W^-)$ but do not show the result on Fig.~\ref{fig:siconst} as the limits are in between these two representative cases. 

In the case of the presence of a thick dark disk we study two extreme scenarios. The first is a fast rotating disk with density $ \rho_{DD}=0.25\rho_x$, lag speed $ | \ v_{\odot} | =150 \, \km / \s$ and velocity dispersion of $\bar{v}_{DD}=86.6 \, \km / \s $. This case is represented by the upper edge of the colored shaded area in Fig.~\ref{fig:siconst}. In the second scenario, we assume a slow rotating disk with $ \rho_{DD}=1.5\rho_x, |\ v_{\odot}| \sim0 \, \km / \s,$ and $ \bar{v}_{DD}=86.6 \, \km / \s$ and it is depicted by the lower edge of the color shaded area in Fig.~\ref{fig:siconst}. The hatched area in Fig.~\ref{fig:siconst} is the region of the parameter space $\sigma_{\chi\chi}/m_\chi < 0.1\, \cm^2 / \g$ for which the self-interaction is not strong enough to resolve the small scale structure problems based on the kinematics of dwarf spheroidals \cite{zavala2013constraining}. 

A dark matter annihilation to heavy leptons potentially has an important consequence. It is possible that for dark matter particle masses between 100 GeV and 10 TeV, self-interaction may be too weak to resolve some of its empirical motivations (e.g., small scale structure) or a dark disk cannot be present in the Milky Way. It is important to emphasize that the thickness of the constraint as shown in Fig.~\ref{fig:siconst} corresponds to the uncertainties in the properties of the dark disk. Nevertheless, even in the most conservative case a self-interacting dark matter of mass ${\cal{O}}(1{\rm{TeV}})$ is inconsistent with a dark disk. 

For completeness in the limit of no self interaction we can apply a similar analysis and observe the effect of the neutrino flux enhancement  by the thick dark disk. Figure~\ref{fig:sisd} shows the constrains obtained assuming spin-independent (SI) interactions  and spin-independent (SD) interactions (where interactions only with Hydrogen were taken into account). For annihilating into heavy quarks we obtain limits only for $m_\chi > 35 \, {\rm{GeV}}$ as that is the lowest energy with constraints from IceCube \cite{aartsen2017search} while for heavy leptons the mass range is $m_\chi > 20 \,{\rm{GeV}}$. The width of the constraints (shaded areas for the two annihilation channels) is set by the range of choices we make on the relative speed and local dark matter density due to the dark disk. This width is smaller at masses $m_\chi < 100$ GeV as compared to $m_\chi > 1$ TeV. This outcome is expected because slower speeds allow for an easier exchange of momentum and energy between the nuclei and the dark matter particle, enough for the latter to be captured into the solar potential well. On the other hand light candidates could have been captured as easily even when they were moving faster and thus the width is narrower in that range. The most constraining power comes at $m_\chi \sim 1$ TeV, and this is set by the IceCube sensitivity curve \cite{aartsen2017search}. 

\section{\label{sec:Intro}Conclusions}

We considered the effects of the presence of a thick dark matter disk on self-interacting dark matter constraints as well as spin-dependent and spin-independent dark matter-nuclei interactions.  A thick dark matter disk is motivated by numerical simulations of the formation of dark matter halos similar to the Milky Way and it is the outcome of the presence of the baryonic disk. The enhanced density and very slow relative velocity between the Sun and the dark matter particles of the disk result in an enhancement of the capture rate of dark matter particles in the Sun. 

We find that the energetic neutrino flux from IceCube limits the self-interacting cross section of dark matter. In the case where the dark matter particle annihilates predominant to heavy leptons we find that it is possible the presence of a dark disk to constrain the self-interaction cross section  to a value less than what is needed to explain small scale structure (one of the motivations for self-interacting dark matter) for dark matter masses between $\sim$ 100GeV and 10 TeV. We also show improved constraints on the interaction between dark matter and nuclei based on the latest IceCube results. 

These results depend on a set of assumptions that if relaxed can have different effects. For example, the value of the relative velocity and velocity dispersion of self-interacting dark matter are obtained from cold dark matter simulations. In reality, if the dark matter possesses a non-zero self-interaction cross section the thick dark disk will evolve over time to a thin but denser disk (e.g., the model of \cite{2013PhRvL.110u1302F, 2014JCAP...06..059F, 2017arXiv170510341R}). Note however that any effect on the structure of the dark disk due to self-interaction will lead to stronger constraints compared to the ones derived here and therefore in the interest of being conservative we did not include such effects. In addition, the results are sensitive to the assumed gravitational potential of the Sun (the escape speed as a function of radius) and internal composition of the Sun as a function of radius (e.g., Table 1 in \cite{1538-4357-705-2-L123} and \cite{2014dapb.book..245B}). 
Both of these quantities have been studied extensively in literature and the effects on the results derived here are negligible compared to the systematics on the distribution of dark matter in the Milky Way halo. 

In summary we show that the hierarchical assembly of the Milky Way places constrains on the self-interacting cross section of dark matter. If future observations of the stellar distribution in the Milky Way gives evidence of the presence of a dark disk in the Milky Way \cite{2014MNRAS.444..515R} it will be possible to further constrain the strength of a possible self-interaction of the dark matter particle. 

\acknowledgements
We acknowledge useful conversations with Victor Debattista, JiJi Fan, George Lake, Justin Read, Darren Reed and Andrew Zentner. This project is funded by DE-SC0017993. SMK thanks the Institute for Computational Science at the University of Z\"{u}rich for hospitality and the organizers of the workshop on Thin, Thick and Dark Disks (Ascona, Switzerland) where this work was conceived. 

\bibliography{manuscript}

\end{document}